\documentclass[acmsmall]{acmart}

\def\markup{0}

\if\markup1
\usepackage[normalem]{ulem}
  \definecolor{myblue}{rgb}{0,0,0.75}
  \newcommand{\rv}[1]{{\leavevmode\color{myblue}#1}}
  \newcommand{\st}[1]{{\sout{#1}}}
\else
  \newcommand{\rv}[1]{#1}
\newcommand{\st}[1]{}
\newcommand{\sout}[1]{}
\fi

\AtBeginDocument{%
  \providecommand\BibTeX{{%
    \normalfont B\kern-0.5em{\scshape i\kern-0.25em b}\kern-0.8em\TeX}}}

\copyrightyear{2024}
\acmYear{2024}
\setcopyright{acmlicensed}\acmConference[CHI '24]{Proceedings of the CHI Conference on Human Factors in Computing Systems}{May 11--16, 2024}{Honolulu, HI, USA}
\acmBooktitle{Proceedings of the CHI Conference on Human Factors in Computing Systems (CHI '24), May 11--16, 2024, Honolulu, HI, USA}
\acmDOI{10.1145/3613904.3642187}
\acmISBN{979-8-4007-0330-0/24/05}

\begin{document}

\title[LightSword: a cognitive training exergame]{LightSword: A Customized Virtual Reality Exergame for Long-Term Cognitive Inhibition Training in Older Adults}

\author{Qiuxin Du}
\orcid{0009-0000-5991-9482}

\affiliation{%
  \institution{Beijing Institute of Technology}
  \city{Beijing}
  \country{China}
}
\email{qiuxin807521@163.com}

\author{Zhen Song}
\authornote{Corresponding Author}
\affiliation{%
  \institution{The Central Academy of Drama}
  \city{Beijing}
  \country{China}
  }
\email{songzhen@zhongxi.cn}

\author{Haiyan Jiang}
\affiliation{%
  \institution{Beijing Institute of Technology}
  \city{Beijing}
  \country{China}
}
\email{jianghybit@163.com}

\author{Xiaoying Wei}
\affiliation{%
 \institution{The Hong Kong University of Science and Technology}
 \city{Hong Kong SAR}
 \country{China}
 }
\email{xweias@connect.ust.hk}

\author{Dongdong Weng}
\affiliation{%
  \institution{Beijing Institute of Technology}
  \city{Beijing}
  \country{China}
  }
\email{crgj@bit.edu.cn}

\author{Mingming Fan}
\authornote{Corresponding Author}
\affiliation{%
  \institution{The Hong Kong University of Science and Technology (Guangzhou)}
  \city{Guangzhou}
  \country{China}
}
\affiliation{%
 \institution{The Hong Kong University of Science and Technology}
 \city{Hong Kong SAR}
 \country{China}
 }
\email{mingmingfan@ust.hk}

\renewcommand{\shortauthors}{Qiuxin and Zhen Song, et al.}

\begin{abstract}
The decline of cognitive inhibition significantly impacts older adults' quality of life and well-being, making it a vital public health problem in today's aging society. Previous research has demonstrated that Virtual reality (VR) exergames have great potential to enhance cognitive inhibition among older adults. However, existing commercial VR exergames were unsuitable for older adults' long-term cognitive training due to the inappropriate cognitive activation paradigm, unnecessary complexity, and unbefitting difficulty levels. To bridge these gaps, we developed a customized VR cognitive training exergame (LightSword) based on Dual-task and Stroop paradigms for long-term cognitive inhibition training among healthy older adults. Subsequently, we conducted an eight-month longitudinal user study with 12 older adults aged 60 years and above to demonstrate the effectiveness of LightSword in improving cognitive inhibition. After the training, the cognitive inhibition abilities of older adults were significantly enhanced, with benefits persisting for 6 months. This result indicated that LightSword has both short-term and long-term effects in enhancing cognitive inhibition. Furthermore, qualitative feedback revealed that older adults exhibited a positive attitude toward long-term training with LightSword, which enhanced their motivation and compliance.
\end{abstract}

\ccsdesc[500]{Human-centered computing~User studies}
\ccsdesc[300]{Human-centered computing~Virtual reality}
\ccsdesc[100]{Human-centered computing~Empirical studies in interaction design}


\keywords{Virtual Reality, Health, Older Adults}

\begin{teaserfigure}
\includegraphics[width=0.9\textwidth]{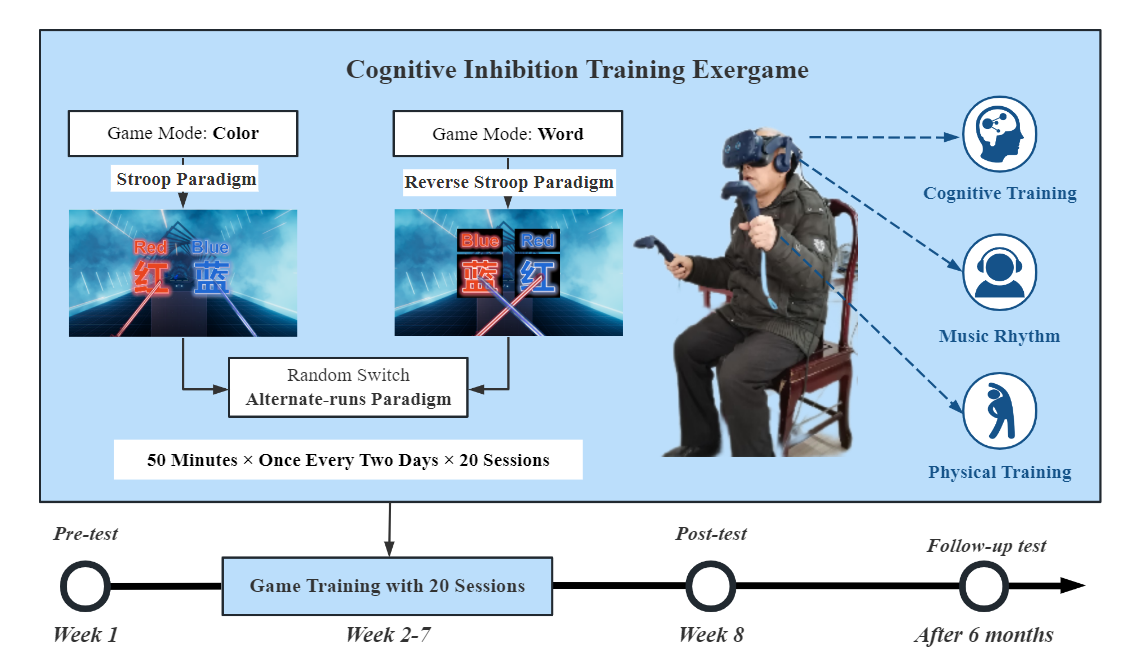}
\centering
  \caption{The figure shows the game design of LightSword. There are two game modes in LightSword: the \emph{\textbf{color}} mode designed based on the Stroop Paradigm, and the \emph{\textbf{word}} mode designed based on the Reverse Stroop Paradigm. The alternating-runs paradigm was used to determine the task order. Subsequently, we conducted a user study to assess the effectiveness of LightSword in enhancing cognitive inhibition function for older adults.}
 \label{fig:teaser}
\end{teaserfigure}

\maketitle

\section{Introduction}


Aging is a natural biological process associated with physical and cognitive decline~\cite{bettio2017effects}. Such decline can negatively affect the quality of life and well-being of older adults~\cite{1,2}. Cognitive decline manifests as deficits in several cognitive processes, including memory, attention, reasoning ability, processing speed, working memory, and executive functions \cite{3,4,5,6}. Without timely intervention, significant declines in cognitive functions can lead to mild cognitive impairment, which may progress to Alzheimer's disease \cite{7,8,46}. The global population is entering an aging phase, and the deterioration of cognitive function could ultimately result in substantial societal and economic consequences  \cite{morawe2012protein, parasuraman1991attention, choi2021ginseng}. Therefore, effective cognitive interventions are imperative and urgent to achieve healthy aging.

Cognitive inhibition is an important component of cognitive function, referring to the ability to suppress ongoing irrelevant information and actions to achieve current goals \cite{14,47,48,49}. However, cognitive inhibition in older adults deteriorates due to structural and functional alterations in the brain \cite{shilling2002individual, tipper1991less}. As a result, they may have difficulties in suppressing and filtering out task-irrelevant information when encoding relevant information into working memory. Such difficulties may overload their limited working memory capacity and reduce their ability to retain and manipulate relevant information \cite{17}. Numerous studies have demonstrated that training in cognitive inhibition can help older adults better filter out distracting information, maintain attention, and reduce cognitive interference, consequently enhancing their cognitive capabilities and overall quality of life \cite{53, mozolic2010cognitive, kuhn2017taking, ji2016plasticity}. Therefore, the preservation and enhancement of cognitive inhibition abilities are of paramount importance in today's aging society.

Encouragingly, abundant evidence suggests that age-related declines in cognitive inhibition can be mitigated or reversed through various approaches, as the adult brain exhibits considerable plasticity \cite{12,53}, encompassing both physical and functional changes. The current mainstream approaches to improving cognitive ability include cognitive training \cite{20,21,22}, physical training \cite{23,24,25}, video game training \cite{26,27,28}, and music therapy \cite{29,30}. Recently, many studies have demonstrated the effectiveness of Virtual Reality (VR) exergames in improving the cognitive inhibition of older adults \cite{41,42,43,44,45,66, huang2020exergaming}. VR exergames combine physical exercise with cognitive training, offering an engaging and appealing training method that motivates older adults to train over extended periods \cite{37,81, cmentowski2023never}. Furthermore, VR exergames could provide rich visual and auditory stimuli. Such immersive experiences keep older adults focused on cognitive tasks and stimulate various cognitive functions of the brain, including attention, memory, problem-solving, and spatial awareness \cite{41,kruse2021enabling,kruse2021long}.


However, there are still substantial gaps that need to be bridged when choosing suitable VR exergames for long-term cognitive training in older adults. Previous studies have demonstrated that commercial VR exergames boost users' motivation with engaging content, clear goals, reward systems, multiplayer modes, and progressively increasing difficulty levels \cite{41,chao2013effects, cmentowski2023never, kruse2022evaluating, sinclair2009exergame}. However, commercial VR exergames also suffer from the disadvantages of inappropriate cognitive activation paradigm, unnecessary complexity, unsuitable difficulty, and susceptibility to cybersickness, which hinder their long-term use as effective cognitive training tools for older adults \cite{44,games2019beat}. Some studies have designed customized VR exergames to mitigate cybersickness or establish dynamic difficulty adaptation~\cite{kruse2021long,44,45}. However, they only conducted short-term experiments to demonstrate the viability and short-term effects of VR exergames, without any long-term studies to explore the long-term effects on cognitive enhancement in older adults \cite{20,58,73}. Little is known about how to design a VR exergame that is suitable for long-term training on cognitive inhibition for older adults. Therefore, we need to develop a customized VR exergame that is not only enjoyable and challenging in long-term training but also tailored to the cognitive and physiological characteristics of older adults.

In this paper, we proposed a VR exergame (LightSword) to conduct long-term cognitive inhibition training for older adults. Considering the age-related decline in cognitive abilities and physical performance in older adults (e.g., slower reaction speed and limited mobility), we made improvements to meet the specific needs of older adults: (1) To ensure the effectiveness of training cognitive inhibitory, we designed LightSword based on the Stroop Paradigm and Dual-task Paradigm \cite{36, 41, 78}. The Stroop Paradigm requires older adults to suppress the interference of irrelevant information: they should ignore words and only identify colors. The Dual-task Paradigm requires older adults to simultaneously accomplish multiple goals: they should not only judge the block correctly but also lift their arms to the target position quickly. \rv{(2) To enhance motivation in encouraging active engagement, customized difficulty levels, real-time feedback and encouragement, individual records, and daily champions have been implemented. (3) To improve compliance for supporting long-term training, we continuously updated the songs and rhythms to provide freshness for older adults while also setting gradually increasing difficulty levels to make the game challenging. (4) To keep safety during prolonged physical exercise, LightSword allowed older adults to play the game while sitting down, with targets flying through the front channel to reduce head movements.} \sout{(2) To practice physical activities with safety, LightSword allowed older adults to play the game while sitting down, with targets flying through the front channel to reduce head movement. (3) To maintain motivation and satisfaction, customized difficulty levels, real-time feedback and encouragement, individual records, and daily champions have been implemented. (4) To support long-term training, we continuously updated the songs and rhythms to provide freshness for older adults while also setting gradually increasing difficulty levels to make the game challenging.}



We conducted a longitudinal user study spanning a total of eight months to demonstrate the effectiveness of LightSword in enhancing cognitive abilities among older adults. Additionally, we discussed how the LightSword's usability, appeal, and feedback mechanism contribute to the overall training experience of older adults. We recruited 12 healthy older adults aged 60 years and above to participate in LightSword training. Each participant was required to complete 20 training sessions over 6 weeks, with each session consisting of 12 songs. Two evaluation methods (quantitative measures and qualitative feedback) were employed to answer our research questions: \textbf{RQ1}: Does LightSword enhance cognitive inhibition in older adults, considering short-term effects and long-term effects? \textbf{RQ2}: How do older adults perceive the experience of LightSword and its usability as a long-term training tool? Quantitative measurements indicated a significant improvement in the cognitive inhibition abilities of the older adults, and this improvement persisted for 6 months. Qualitative feedback revealed that older adults expressed a positive attitude toward long-term training with LightSword. To our knowledge, we were the first to use a custom-designed VR exergame for long-term cognitive inhibition training in older adults.

The contributions of this paper are:
\begin{itemize}
\item We proposed a customized VR exergame for long-term training among older adults and key elements for aging-appropriate game design.
\item A longitudinal user study spanning a total of eight months was conducted to explore the usability of LightSword and its effectiveness in improving older adults' cognitive inhibition.
\item The results demonstrated the potential of LightSword to improve cognitive inhibition performance in older adults, particularly in short-term and long-term effects. Meanwhile, older adults appreciated LightSword and expressed a willingness to receive long-term training.
\end{itemize}

\section{Related Work}

\rv{To clarify our research motivation and research gaps, we reviewed related work about cognitive training, focusing on four aspects: 1) We reviewed the cognitive inhibition in older adults and the severe consequences of such decline; 2) We outlined the evolution of cognitive training games and found that games based on the cognitive activation paradigm are highly popular: 3) We summarized various cognitive activation paradigms used for cognitive inhibition training; 4) We focused on the advantages and disadvantages of existing VR exergames tailored for older adults.} \par
\sout{In this section, we first review the older adults' cognitive inhibition. Then, we focus on the current games designed to provide cognitive intervention for older adults.}

\subsection{Cognitive Inhibition in Aging}


Cognitive inhibition means that individuals consciously suppress or control certain cognitive processes, responses, or impulses to better accomplish the current task or objective when processing information or performing tasks \cite{houde2000shifting}. \rv{Previous research has demonstrated that cognitive inhibition processing capacity tends to decline with age, significantly impacting the daily functioning and independence of older adults, and potentially causing irreversible brain health damage \cite{hasher1988working}. Mild cognitive inhibition impairment manifests as attention difficulties, increased distractibility, memory decline, and challenges in expression, particularly in complex or distracting situations. If not intervened timely, this impairment could lead to severe consequences. An increasing number of research studies have associated cognitive inhibition deficits with elderly depression. Joormann et al. have highlighted cognitive inhibition as a crucial mechanism in emotion regulation \cite{joormann2007cognitive}. S. Richard-Devantoy's finding suggested that the inability to inhibit neutral information access to working memory, restrain and delete irrelevant information may impair the patient's capacity to respond adequately to stressful situations subsequently leading to an increased risk of suicidal behavior during late-life depression \cite{richard2012deficit}. Therefore, enhancing cognitive inhibition in older adults is a serious public health concern, and finding effective intervention measures is highly necessary and urgent.} \par

\sout{Due to the plasticity of the human brain, older adults have the potential to maintain and enhance their cognitive inhibition abilities through training. For example, Anguera et al. designed a driving game based on the dual-task paradigm for older adults to evaluate and train their cognitive control abilities. The results suggested that video games serve as a powerful tool for cognitive enhancement. They also demonstrated the substantial potential of game design based on cognitive activation paradigms for cognitive interventions \cite{26}. Previous research indicated that the processing capacity of cognitive inhibition tends to deteriorate with age, making older adults more susceptible to distraction compared to younger adults \cite{hasher1988working}. An increasing number of studies are opting to design games based on cognitive activation paradigms for cognitive inhibition training in older adults. Zhang et al. developed the exergame (Ping Pong) incorporating the traditional Go/NoGo paradigm which requires the participant to perform an action only when given certain stimuli. The results showed that the Ping Pong exergame received a good usability score and participants presented significantly better performance in cognitive inhibition tasks after training \cite{53}. With the development of VR, some researchers explore the cognitive inhibition enhancement effects of combining VR with psychological paradigms. Parsons et al. compared the differences between the VR Stroop task (VRST) and the paper-and-pencil Stroop task in enhancing supervised attentional processes. The findings showed VRST has a greater advantage in enhancing cognitive inhibitory \cite{66}. Therefore, VR games designed based on psychological paradigms are an effective intervention for improving cognitive inhibition in older adults.}

\subsection{\rv{Games for Cognitive Inhibition Training}}
\rv{Due to the plasticity of the human brain, older adults have the potential to maintain and enhance their cognitive inhibition abilities through training. With the development of information and communication technology (ICT), game training is gradually being incorporated into cognitive training, and the platforms used: computers, smartphones, tablets, and the latest head-mounted displays—are always being updated. Video game-based cognitive interventions may be ideal for combating the many perceptual and cognitive declines associated with advancing age. Game interventions have the potential to induce higher rates of intervention compliance compared to other cognitive interventions as they are assumed to be inherently enjoyable and motivating. Anguera et al. designed a driving game based on the dual-task paradigm for older adults to evaluate and train their cognitive control abilities. The results suggested that video games serve as a powerful tool for cognitive enhancement. They also demonstrated the substantial potential of game design based on cognitive activation paradigms for cognitive interventions \cite{26}. An increasing number of studies are opting to design games based on cognitive activation paradigms for cognitive inhibition training in older adults. Zhang et al. developed the exergame (Ping Pong) incorporating the traditional Go/NoGo paradigm which requires the participant to perform an action only when given certain stimuli. The results showed that the Ping Pong exergame received a good usability score and participants presented significantly better performance in cognitive inhibition tasks after training \cite{53}. With the development of VR, some researchers are eager to explore the cognitive enhancement effects of combining VR with cognitive activation paradigms. Parsons et al. compared the differences between the VR Stroop task (VRST) and the paper-and-pencil Stroop task in enhancing supervised attentional processes. The findings indicated that VRST has a greater advantage in enhancing cognitive inhibitory \cite{66}. Therefore, VR games designed based on cognitive activation paradigms are an effective intervention for improving cognitive inhibition in older adults.}

\subsection{\rv{Cognitive Activation Paradigms for Cognitive Inhibition}}
\rv{The cognitive activation paradigm refers to the utilization of specific cognitive tasks or activities to activate particular cognitive functions or neural processes. It is commonly employed in the design of cognitive training, neurofeedback, and intervention programs. Cognitive game training based on cognitive activation paradigms is an effective tool for enhancing cognitive abilities. They provide targeted training and cognitive stimulation for corresponding brain regions, while also enhancing motivation and engagement. There are many cognitive activation paradigms for cognitive inhibition, such as the Stroop task \cite{60}, Reversed Stroop task \cite{61,62,63}, and Flanker task \cite{64,65}, the Simon task \cite{32,33} and Go/NoGo task \cite{30}. Parsons et al. evaluated the VR Stroop task (VRST) in enhancing supervised attentional processes, and found VRST is more conducive to increased inhibitory control, and VRST's complexity condition may be used to evaluate external and endogenous attention \cite{29}. Wang et al. found that after three weeks of inhibition training with immediate feedback, older adults showed a transfer effect on the Go/NoGo task, providing evidence that inhibition training may have a transfer effect \cite{35}. Furthermore, the dual-task paradigm and task-switching paradigm are frequently employed in cognitive training as auxiliary tools within game tasks. Strobach et al. showed that participants gained a performance advantage when completing two activities simultaneously, suggesting that switching between dual activities can improve executive control abilities \cite{30,66,67}. The task-switching paradigm revealed that participants had significantly higher reaction times when completing switching tasks compared to repetitive tasks, indicating the presence of switch costs \cite{31}. Therefore, choosing the appropriate cognitive activation paradigm plays a crucial role in the effectiveness of cognitive training games.} \par

\subsection{\rv{Customized VR Exergames for Older Adults}}

In recent years, VR exergames have become popular in cognitive interventions due to their ability to simultaneously engage the physical and cognitive training \cite{35}. A recent study reported favorable outcomes in terms of usability and positive experiences among our older adult participants when using commercial VR games, such as BoxVR, during the pandemic  \cite{boxvr, campo2021immersive}. Some studies also demonstrated that commercial VR exergames are superior in terms of playfulness, encouraging motivation, and real-time feedback \cite{kivela2019study, palaniappan2018developing}. However, they were not suitable for older adults for long-term training \cite{44, cmentowski2023never}. For example, Kruse et al. compared a VR exergame (Maestro Game VR) to a video-guided exercise in older adults. Their findings suggested that Maestro Game VR performed better on playfulness and motivation, but there are still challenges in terms of cybersickness and the repetition and boredom of long-term training \cite{kruse2021enabling}. Furthermore, commercial games also had some disadvantages of unnecessary complexity, and inappropriate difficulty \cite{44,games2019beat}. So customized VR exergame designs were increasingly being embraced to further enhance the effectiveness of cognitive training. Lucie et al. developed a motor-cognitive exergame (Memory Journalist VR) for people with dementia using a human-centered design approach \cite{kruse2021long}. The findings indicated that custom-designed exergame can help to improve and maintain cognitive, physical, and psychological well-being. Sukran et al. designed and implemented a VR exergame (Canoe VR) for older adults. They highlighted VR technology is still novel and exposes some difficulties for users such as usability problems, cybersickness, or ergonomics \cite{44}. \rv{What's more, current existing VR exergames did not take into account the compliance of long-term training. Boot et al. found that the exergame, though presumed to be more beneficial in previous research, resulted in the lowest compliance \cite{boot2013video}. Some studies also emphasized the importance of considering age-related cognitive differences and changes in physical movements when designing games \cite{abeele2021immersive, hirzle2021critical, laviola2000discussion}. } \par
\rv{Overall, when designing VR exergames specifically tailored for long-term cognitive training in older adults, there are several research gaps that need to be bridged: 1) the personal cognitive and limited mobility characteristics of older adults; 2) the motivation and compliance to support long-term training; 3) the safety and cybersickness in VR exergames. In our work, we try to answer the question: how can we design a VR exergame to support long-term cognitive training among older adults?}

\section{Design Goal}
\rv{Based on the insights from prior literature and the feedback of older adults, we designed a VR exergame (LightSword) using the Unity game engine to provide an engaging and effective way to train older adults' cognitive inhibition \cite{unity}. Our primary focus was on improving the participants' motivation and compliance to support long-term training. Equally important is considering the needs of older adults, like declining cognition, slower reaction times, and limited mobility, and making them feel satisfied and challenged. Additionally, we consider the safety and comfort of physical activities for older adults.} \par

In this section, we outline four design goals (DGs) to support older adults' needs and long-term training, \rv{including Effectiveness, Motivation, Compliance, Safety.}


\subsection{\rv{DG1: Effectiveness of Training Cognitive Inhibition}}

\rv{Cognitive inhibition is crucial for older adults as it involves abilities such as impulse control, ignoring distractions, and sustaining focus. Maintaining a strong cognitive inhibition capacity benefits the cognitive health and daily functioning of older adults \cite{14,16,18}. Therefore, our study focuses on cognitive inhibition training.} Cognitive training aims to enhance an individual's cognitive abilities and intellectual functions through a series of specific cognitive tasks, exercises, and activities \cite{hertzog2008enrichment, ball2002effects}. Much research has demonstrated that incorporating cognitive activation paradigms into games is a highly effective intervention \cite{26, 53}. Different cognitive training relies on distinct brain regions and requires different activation paradigms. \rv{Previous research suggested that training gains are often extremely specific \cite{boot2011mental,boot2013video}.} Therefore, we believe that the choice of cognitive activation paradigms is important for successful cognitive training. \rv{A suitable cognitive activation paradigm targeting cognitive inhibition may lead to more lasting effects in long-term training \cite{hung2018dissociations, smith2009correspondence, martin2010paradigms}}. In the prototype design, the most important goal is to select appropriate psychological paradigms for training inhibitory control, integrating them seamlessly into our game tasks.


\subsection{\rv{DG2: Motivation in Encouraging Active Engagement}}

Motivation and satisfaction are key factors in determining the effectiveness of gamified training \cite{gizatdinova2022pigscape}. When participants experience both interest and challenge in a game, they enter the flow state, leading to increased engagement and greater time and effort invested \cite{84}. \rv{Therefore, it is necessary to establish an appropriate level of game difficulty in game design because excessive complexity can reduce the enthusiasm of older adults, while excessive simplicity can lead to boredom \cite{alexander2013investigation, zohaib2018dynamic}. 
To establish an appropriate level of game difficulty and maintain the motivation of older adults, we should consider the cognitive and physiological characteristics of older adults, such as age-related cognitive and physical declines, slower reaction times, and limited mobility issues.} \sout{When participants find a game both interesting and challenging, they are more likely to engage actively and invest more time and effort in playing the game. Specifically, the game should be easy to learn, extremely interesting and highly usable, even for beginners. Striking the right balance in-game difficulty is crucial, as excessive complexity may dampen the motivation of older adults, while excessive simplicity can lead to boredom. To maintain motivation among older adults, it is imperative to adjust the difficulty levels, considering age-related declines in cognitive and physical abilities, such as slower reaction times and mobility issues, during the game design process.} Additionally, in-game social interaction offers numerous benefits for older adults, including reducing feelings of loneliness, enhancing emotional well-being, fostering social engagement, providing cognitive stimulation, and encouraging sustained game participation \cite{ristau2011people, nezlek2002psychological, lee1987social}. Furthermore, real-time feedback and encouragement could provide a sense of achievement and fulfillment in older adults \cite{gizatdinova2022pigscape}. All these elements should be seamlessly incorporated into our game design to motivate older adults.

\subsection{\rv{DG3: Compliance to Support Long-term Training}}

\rv{Participants' compliance, defined as their continuous participation and adherence in therapy or training programs, is crucial for cognitive training. Higher compliance indicates more active engagement in the training, often correlating with better training outcomes \cite{ice1985long, winnick2005you}. Particularly in long-term training programs, maintaining high compliance is essential for achieving training objectives and maximizing effectiveness. Therefore, we believe that supporting long-term training is crucial for VR cognitive exergames, as older adults require sustained improvements in cognitive domains to have a broader positive impact in their daily lives, beyond short-term performance enhancements \cite{20, gunther2003long}.} \par

\sout{ Supporting long-term training is crucial for VR cognitive exergames, as older adults require sustained improvements in cognitive domains to have a broader positive impact in their daily lives, beyond short-term performance enhancements. Therefore, we should enhance the potential for long-term training in older adults through gradually increasing difficulty levels, constantly updated game content and a degree of competition. }

\subsection{\rv{DG4: Safety in Prolonged Physical Exercise}}

Exergame refers to an entertainment activity that integrates physical exercise and video games, allowing participants to engage in physical activity in an enjoyable way \cite{37}. Some intense physical activities such as jumping, dodging, or moving quickly may pose physical challenges, increasing the risk of falls or injuries \cite{weech2019presence, laviola2000discussion}. \rv{Therefore, when designing exergames for older adults, we should strive to avoid such intense movements.} What's more, older adults are particularly susceptible to suffering these risks and discomfort during VR exergame due to age-related declines in their balance and proprioceptive abilities, reduced adaptability to exercise, and limited exposure to VR devices \cite{drazich2023too, arns2005relationship, dilanchian2021pilot, wei2023bridging}. Therefore, during the game prototype design phase, it is essential not only to ensure the effectiveness of physical exercise but also to reduce cybersickness and potential security risks. \par

\section{Game Implementation}

The goal of the LightSword exergame is to provide a fun and engaging way to train older adults' cognitive inhibition. The game scene is inspired by Beat Saber \cite{games2019beat}. In our game, blocks fly from two channels following the rhythm of the music. Older adults hit the targets by swinging the HTC VIVE controllers, which are mapped as two lightsabers in the virtual scene. Older adults hold a red lightsaber in their left hand and a blue lightsaber in their right hand. When the lightsaber successfully hits the block, they earn a corresponding score. \par

In this section, we describe the technical implementations and the last version of the game prototype according to design goals.

\begin{figure}[h]
\centering
\includegraphics[width=0.9\linewidth]{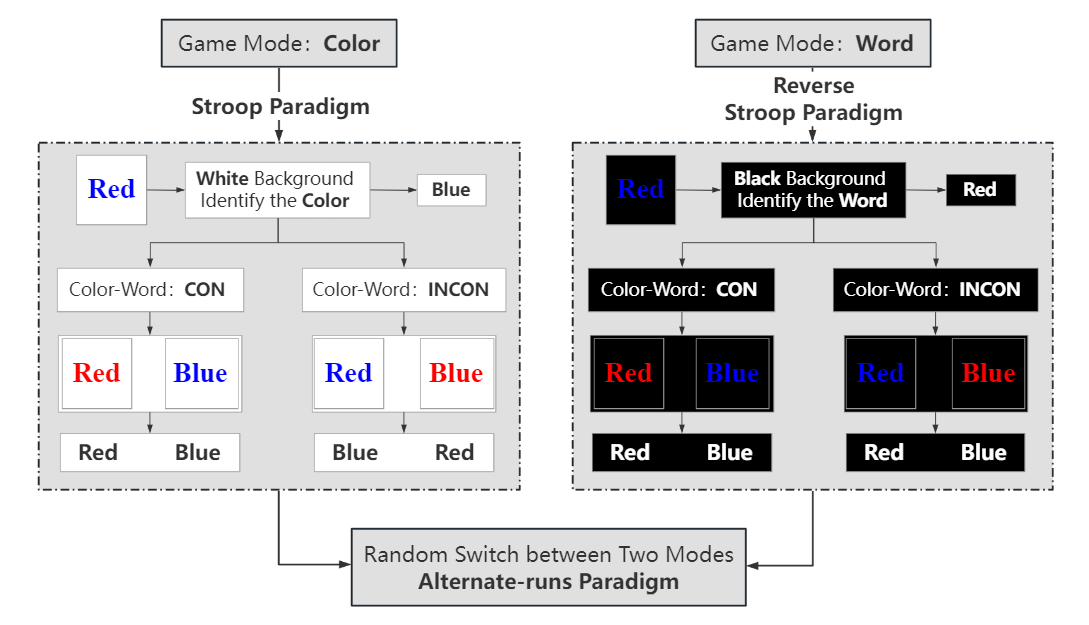}
\caption{The specific design ideas and methods of two game modes}
\label{Fig:mode}
\end{figure}

\subsection{Select Appropriate Cognitive Activation Paradigms to Train Cognitive Inhibition}

To train cognitive inhibitory (\textbf{DG1}), we chose the Stroop paradigm and the Reverse Stroop paradigm. \rv{Because this paradigm, combined with the game, was user-friendly for beginners, required lower educational levels and visual capabilities for older adults, and was easily operable in VR.} Numerous studies have demonstrated the effectiveness of Stroop for enhancing cognitive inhibition, and the Stroop paradigm was widely accepted as one of the activation paradigms for cognitive inhibition \cite{36,41}. When the Stroop effect occurred, the individual's selective attention and conflict control mechanisms were simultaneously engaged. We designed two modes as shown in \autoref{Fig:mode}: \textbf{\emph{color}} mode was based on the Stroop paradigm, utilizing a white background to identify the color and overcome word interference, while the \textbf{\emph{word}} mode was based on the Reverse Stroop paradigm, employing a black background to distinguish the word and overcome color interference. Additionally, we incorporated the alternate-runs paradigm (It refers to an individual's cognitive flexibility and switching ability when dealing with multiple tasks or stimuli) into the game playing, which may improve cognitive flexibility and cognitive inhibition in older adults \cite{15, ionescu2012exploring}. It required older adults to remember not only the rules of \textbf{\emph{color}} mode and \textbf{\emph{word}} mode, but also remember the sequence of the two modes. The dual-task paradigm refers to individuals simultaneously performing two tasks, and it can enhance multitasking and cognitive inhibition abilities \cite{78}. Therefore, we added the dual-task paradigm in the game, which required older adults to quickly identify the block while simultaneously swinging the controller to hit the target. \rv{We naturally integrated various cognitive activation paradigms into the block design, providing richer cognitive stimulation and enhancing the diversity and amusement of the game. Importantly, this design demanded higher cognitive costs without requiring older adults to master more complex operations.}


\subsection{Customized Settings to Enhancing Motivation in Encouraging Active Engagement}

To achieve \textbf{DG2}, we applied the following design strategies.

\subsubsection{Customized Gradually Increasing Difficulty} 
\
\newline
To adjust the game difficulty to suit older adults, we conducted assessments on five older adults without Mild Cognitive Impairment (MCI) before system implementation \cite{58}. Their performance in the Stroop task was measured using Eprime \cite{eprime}, revealing an average reaction time of 756 milliseconds (ms). The mean longest reaction time was 1423ms and the mean shortest reaction time was 556ms. As the proportion of inconsistent trials increased, the accuracy of older adults gradually decreased. According to the reaction time of older adults and the accuracy of inconsistent blocks, we established three customized progressively increasing game difficulty levels. In the Easy level, the proportion of inconsistent blocks was between 20$\%$ and 30$\%$, the distance between neighboring blocks was more than 60cm, and the block flight time was more than 1.5 seconds; in the Normal level, the proportions were at 30$\%$-40$\%$, distances at 50-60cm, and flight times at 1-1.5 seconds; and in the Hard level, the proportions were at 40$\%$-50$\%$, distances at 40-60cm, and flight times at 0.5-1 seconds. Flight speed varied in range with the rhythm of the songs.

\subsubsection{Real-time Feedback and Encouragement} 
\
\newline
We incorporated algorithms to calculate and display scores in real-time. This real-time feedback could allow participants to view their scores when they correctly hit a block, along with tracking the number of consecutive hits and the overall game score, all conveniently displayed on the left side of the scene. As a reward for achieving five consecutive ``GREAT'' scores, an ``Excellent'' message along with bubbles will appear from the sides of the scene, as shown in \autoref{fig:sence}. This real-time feedback mechanism provided positive reinforcement, elevating their motivation and fostering a sense of accomplishment. 

\subsubsection{Individual Records and Daily Champion}
\begin{figure}[h]
\centering
\includegraphics[width=\linewidth]{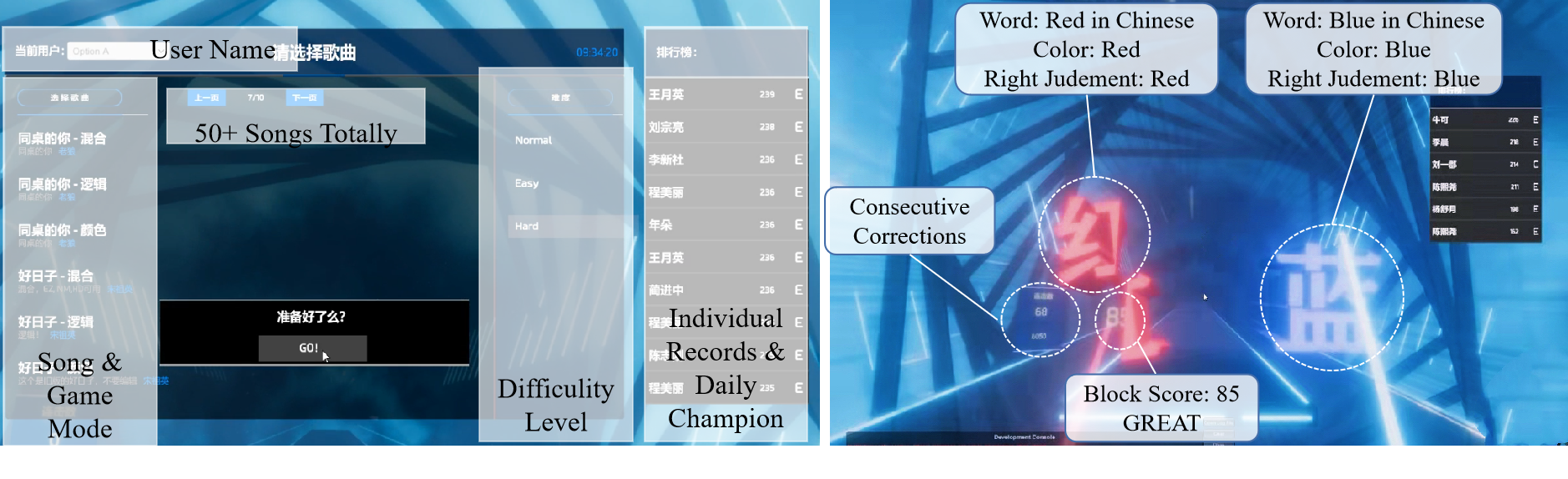}
\caption{Virtual Environment of LightSword: Left: the UI components of the game; Right: Participants are training with LightSword's color mode.}
\label{fig:sence}
\end{figure}

\
\newline
In the game, we offered straightforward and appealing components, complemented by user-friendly prompts, such as participants, modes, and difficulty levels, as well as a leaderboard to show individual records and daily champions, as shown in \autoref{fig:sence}. Additionally, we integrated algorithms for real-time calculations and ranking updates within the game, \rv{to strengthen social connections and maintain competitiveness}. Older adults could review their historical records and today's winner after the completion of a song. Achieving a higher score in the next song will result in an immediate advancement in the rankings. The daily winner will receive milk as a reward.

\subsection{\rv{Maintain Freshness by Updating Songs to Improve Compliance and Support Long-term Training}}

Utilizing music associated with participants' personal biographies could evoke memories and create a soothing ambiance \cite{29,30,33}. To achieve \textbf{DG3}, we interviewed older adults aged 60 years and above to collect songs they liked and were familiar with. Eventually, we carefully selected more than 50 songs that older adults could sing along with. Different songs contained unique musical elements such as rhythm, beat, melody, and emotion, which contributed to the diversity of the game, allowing participants to experience varying challenges and enjoyment across different songs. This was also one of the reasons why BeatSaber gained widespread popularity \cite{shultz2017music, begel2017music}. By changing different songs, we could create varied game experiences. \rv{As the difficulty increased, we opted for faster-paced music accompanied by more frequent appearances of blocks and higher arm movement demands.} This mechanism effectively reduced boredom and aversion among older adults during repetitive training.

\begin{figure*}[tb]
\centering 
\includegraphics[width=0.8\textwidth]{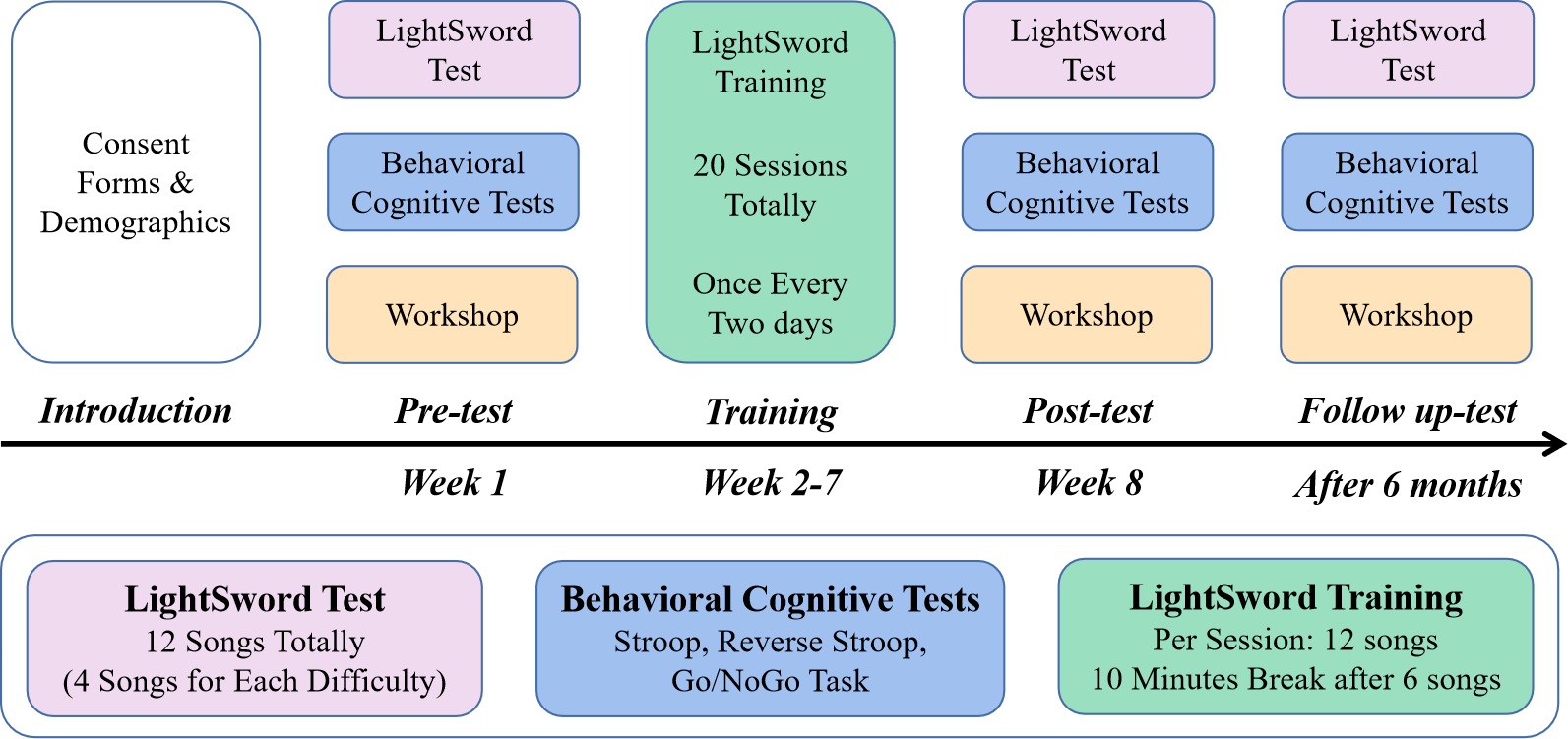}
\caption{Experimental procedure. The below part is an additional description of the above part.}
\label{fig:procedure}
\end{figure*}

\subsection{\rv{Reduce Head Movement and Remain Seated to Ensure Safety in Prolonged Physical Exercise}}

Large-scale movements of the head and body can cause cybersickness and safety risks, which become more severe as the duration of VR exposure increases. In our game, target blocks fly from two channels in front of both eyes, eliminating the need for participants to rotate their heads to observe the targets. Older participants could interact with the targets by waving their arms while remaining seated, which could reduce the risk of falling or getting injured due to intense physical activity or cybersickness. To achieve \textbf{DG4}, we customized the score for each target block to ensure the effectiveness of physical training. The scoring mechanism for each block was as follows: correctly identifying the target yields 60 points, hitting it with speed earns 25 points (requiring participants to swing their arms quickly), and hitting it within the designated area (requiring participants to raise their arms to the target position) earns 15 points. If the score exceeds 60 points, it is recorded as ``GREAT'', but if a hit is missed, it is recorded as ``MISS''.

\section{Methods}

We conducted a longitudinal user study to demonstrate the effectiveness of LightSword in enhancing cognitive inhibition among older adults. Additionally, we were interested in how the game's usability, appeal, and feedback contribute to older adults' overall game experience. Considering these motivations for our user study, we employed two evaluation methods: Quantitative Measures and Qualitative Feedback, to answer our research questions.


\subsection{Participants}
\rv{We calculated the total sample size using G*Power 3.1 \cite{gpower}. For this study, we used the repeated measures ANOVA, at a significance level ($\alpha=0.05$) and large effect size ($f=0.4$), the predicted total sample size to achieve an 80$\%$ statistical power is at least 35 for three repeated measures. Therefore, we recruited a total of 12 participants for our study.} \par

\rv{Based on the United Nations standards for the older adults and the local standards in the HCI community \cite{caine2016local, older},} We recruited 12 healthy older adults aged \rv{60 years and above} \sout{60 to 80} (mean age 65.68±3.89 years, 1 male and 11 females) for our study. All participants met the inclusion criteria, including having no significant medical illnesses, no history of mental or neurological abnormalities, and no sensory impairments. Their vision and hearing capacities were normalized. To ensure the mental well-being and independent living ability of the older adults, they completed several questionnaires including the Mini-Mental State Examination (MMSE), Geriatric Depression Scale (GDS), Activity of Daily Living Scale (ADL), and Self-rating Anxiety Scale (SAS). All participants' scores fell within the healthy range, indicating their suitability as study subjects. Before the start of the experiment, written informed consent was obtained from each participant, with the experimental purpose and task. All participants were informed of their right to withdraw from the study at any time, although no participants chose to do so. All participants were given payment at the end of the experiment. The entire study was conducted in a hospital and received approval from our university's ethics and morals committee. \rv{It was worth noting that the participants aren't isolated within the hospital, which simply provides us with a controlled and hygienic environment for experiments during the pandemic.}

\begin{figure*}[tb]
\centering 
\includegraphics[width=\textwidth]{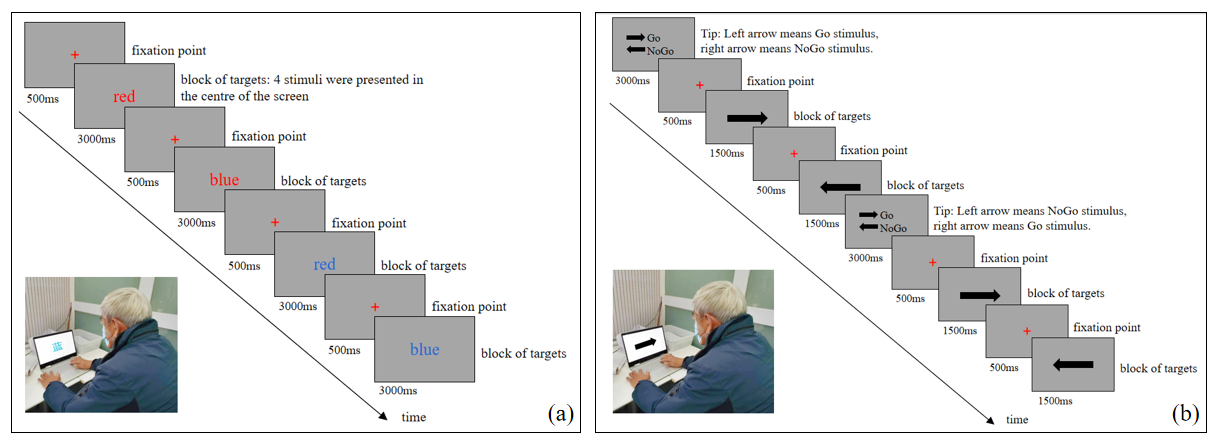}
\caption{The figure shows the procedure of behavioral cognitive test: (a) Stroop and Reverse Stroop tasks, (b) Go/NoGo task.}
\label{fig: behavioral}
\end{figure*}

\subsection{Experimental Procedure}

The experimental procedure is shown in \autoref{fig:procedure}.
Before the start of the experiment, all participants were required to complete an informed consent form and a questionnaire which was used to gather demographic information and assess their familiarity with VR games. Subsequently, the pre-test phase commenced, where all participants were invited to perform one game test (comprising a total of 12 songs, with 4 songs for each difficulty level) and three behavioral cognitive tests (The Stroop task, the Reversed Stroop task, the Go/NoGo task). To minimize the testing effect, participants were instructed to practice many times until they fully understood the testing rules before the formal testing commenced. After the test, we conducted a workshop to gather feedback from the participants regarding the game. \par

The training phase spanned from Week 2 to Week 7. Each participant was invited to engage in a total of 20 sessions of LightSword with three different difficulty levels within 6 weeks, divided into 7 sessions for the easy level, 7 for the normal level, and 6 for the hard level. Each training session comprised 12 songs. The game mode for each session was randomly assigned by the experimenter and did not follow a specific order. After training on 6 songs, the older participants were allowed a 10-minute break. Each training session took approximately 40-50 minutes. Experimenters were present throughout the training to provide necessary assistance. \par

Upon completing the game training phase at week 8, older participants entered the post-test phase, where they were invited to perform a game test and three behavioral cognitive tests (the same as the pre-test). Finally, older participants were asked to complete a workshop to gather long-term use feedback on the LightSword. \par

After 6 months, we conducted the follow-up test, where participants underwent a game test and three behavioral cognitive tests (the same as the pre-test) once again. 

\subsection{Quantitative Data Collection and Analysis}

We employed quantitative measures to demonstrate the effectiveness of LightSword in improving cognitive inhibition in older adults. Primary measures included: short-term effects and long-term effects. Short-term effects refer to the outcomes observed immediately or within a brief period following cognitive training. Long-term effects denote the sustained impact of cognitive training, observed over several months or even years \cite{20,58,73}. To prevent experience effects on game scores, we added three behavioral cognitive tests to validate RQ1.

\subsubsection{Game Scores}
\
\newline
Based on the calculation method of the Stroop paradigm (accuracy and response time) and our customized design, considering the amplitude and speed of the older adults' arm movements, we used the following algorithm to calculate the game score (GS):
\begin{equation}
GS=\frac{cor}{total}*2-\frac{fault}{total}-\frac{miss}{total}+\frac{great}{total}*3
\end{equation}
\par
Cor: the number of correct blocks, \rv{which reflects the ability of older adults to overcome distractions and correctly judge targets}; Fault: the number of fault blocks; 
Miss: the number of missing blocks;
Great: the number of great blocks, \rv{which is difficult for older adults because it requires not only accuracy but also the speed and height of the lifted arms.} \par
\rv{The reason for this equation lies in our emphasis on the accuracy of older adults' gameplay. This emphasis serves as a partial reflection of their cognitive inhibition, which relates to their capacity to successfully overcome distractions and make accurate judgments regarding targets. Simultaneously, we highlight the progress in the GREAT rate, which imposes constraints on the speed and height of arm movements, providing insights into the physical coordination and reaction time of older adults. It is worth noting that attaining a high score in GREAT can be challenging for older adults.}

\subsubsection{Behavioral Cognitive Tests}
\
\newline
We adopted three behavioral cognitive tests to evaluate older adults' cognitive inhibition in this study. 

\begin{itemize}
    \item \emph{The Stroop task}: The Stroop task is commonly used to assess cognitive inhibition \cite{66}. Participants in the Stroop task were required to identify the color while overcoming word interference. The experimental materials consisted of two different colors of Chinese characters, namely ``red'' and ``blue'', with a total of four stimulus materials. These materials included two materials with congruent color-word (CON) and two materials with incongruent color-word (INCON). Each trial followed a specific sequence: a red gaze point ``+'' appeared at the center of the screen for 500 ms, followed by a 3000 ms presentation of the stimulus, as shown in \autoref{fig: behavioral}. Participants were instructed to respond as quickly as possible by pressing the ``Q'' key on the keyboard for red characters and the ``P'' key for blue characters. The experiment consisted of 80 trials for each condition: color-word congruent (20 trials per material * 4 materials) and color-word incongruent (20 trials per material * 4 materials). The Stroop interference effect, calculated by subtracting accuracy (ACC) and response time (RT) between the congruent and incongruent trials, was used to assess participants' conflict inhibition abilities. A smaller Stroop interference effect indicated a stronger conflict inhibition capacity.
    
    
    \item \emph{The Reversed Stroop task}: The Reversed Stroop task is also used to assess cognitive inhibition \cite{67}. Participants in the Reversed Stroop task should make discrimination based on the word of the block and overcome color interference. Other conditions were completely the same as in the Stroop task.

    \item \emph{The Go/NoGo task}: The Go/NoGo task is commonly used to assess response inhibition \cite{klein1997effect,nosek2001go}. Two separate stimuli were used in the experiment: the left arrow and the right arrow. Each trial followed a specific sequence: A 500-ms red gaze point ``+'' was shown first, then a 500-ms stimulus (the Go and NoGo stimuli displayed randomly in the center of the screen), a 1500-ms response page after the stimulus disappeared, and finally a 500-ms random blank screen, as shown in \autoref{fig: behavioral}. The participant quickly pressed the space key on the keyboard when the Go stimulus disappeared and the response page appeared, while the participant did not reply when the NoGo stimulus occurred. The formal experiment was divided into two blocks: one with a left-arrow Go stimulus and a right-arrow NoGo stimulus, and the other was the opposite with a right-arrow Go stimulus and a left-arrow NoGo stimulus. For a total of 160 trials, 80 NoGo trials, and 80 Go trials (1:1). The discriminability index d, and Go response time (RT) were used to assess the participants' response inhibition ability. Greater inhibition was indicated by higher d values. $d'=z(hit \;rete)-z(false \; alarm \; rate) $ . The hit rate is the probability of a participant responding correctly to a Go stimulus; the false alarm rate is the probability of a participant responding incorrectly to a NoGo stimulus, which should stop responding.
    
\end{itemize}

\subsection{Qualitative Data Collection and Analysis}

We gathered qualitative feedback from the older adults through open-ended questions in the workshop to understand their experience with LightSword. In these questions, we particularly focused on four topics: \textbf{Benefits from Exergame Training, Motivation and Compliance, Safety and Usability, User Preferences}. We used the following open-ended questions to capture the perspectives of the older adults on these aspects:

\begin{itemize}
    \item \emph{\rv{``What did you perceive as the benefits from exergame training, and why?''}}
    \item \emph{\rv{``Which part of LightSword could support your motivation and long-term training, and why? How long would you like to train? ''}}
    \item \emph{``Did you experience any issues related to safety or usability during your gameplay in VR? If so, why do you think they occurred?''}
    \item \emph{``Which aspect of the game would you like to see changed or retained, and why?''}
\end{itemize}

\subsection{Apparatuses}
Our training system was built on a computer and older adults use the Head-Mounted Display (HMD) and a handheld controller for game training. We used NVIDIA GeForce GTX 1080 Ti and paired it with Intel Core i7-8700. The game graphics were displayed on a Lenovo LXGJ556D monitor, a 21-color display with a resolution of 1024*768, and a refresh rate of 85 Hz. The HMD was HTC VIVE Pro2, with a monocular resolution of 2448*2448, a refresh rate of 120 Hz, and a maximum field of view of 120 degrees. Behavioral cognitive tests were conducted on a computer, with older adults responding via the keyboard. E-prime was used to run the experiment, and the stimulus presentation and response times were generated automatically by the computer.

\section{Results}
\subsection{Quantitative Results}

We employed quantitative measures to demonstrate the effectiveness of LightSword in improving cognitive inhibition in older adults. To prevent the influence of experience on game scores, we incorporated additional behavioral cognitive tests.


\subsubsection{Game Scores}
\begin{figure}[htbp]
\centering 
\includegraphics[width=3 in]{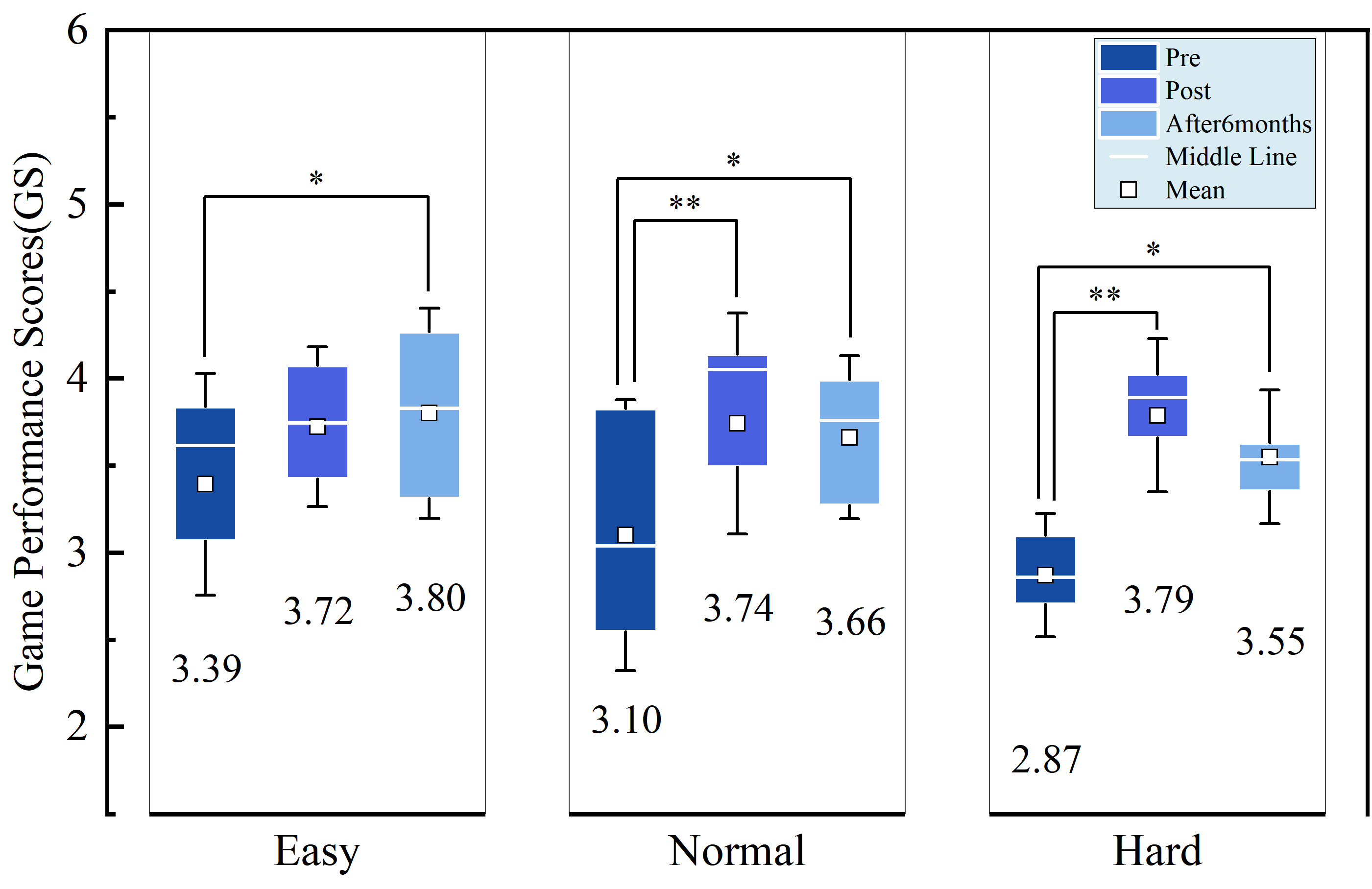}
\caption{The GS of older adults in the pre-test, post-test, and after 6 months-test. ($\**: p<0.05$, $\***: p<0.01$, $\****: p<0.001$)}
\label{fig: game time}
\end{figure}

\
\newline
To assess the differences before and after training, we conducted a two-way repeated-measures ANOVA with 3-time points (pre, post, and 6 months after training) * 3 game difficulties (easy, normal, and hard), as shown in \autoref{fig: game time}. The results revealed a significant main effect of time $(F_{2,66}=13.700, p=0.000)$. Post-hoc analysis indicated that the GS of older adults was significantly higher in the post-test compared to the pre-test $(p=0.000)$, and these improvements remained significant 6 months later $(p=0.003)$. However, there was no significant difference in the participants' GS as the difficulty level increased $ (F_{2,33}=0.307, p=0.738)$. Furthermore, the interaction effect between time and difficulty did not reach statistical significance $ (F_{4,66}=2.416, p=0.057)$. \par
In addition, we conducted separate one-way analysis of variance (ANOVA) tests for the three levels of difficulty in the pre-test. The results indicated significant differences in game scores among older adults across different difficulty levels $(F_{2,33}=6.006, p=0.032)$. Post-hoc comparisons revealed that game scores in the easy difficulty level were significantly higher than those in the Normal $(p=0.043)$ and Hard $(p=0.013)$. \par

\begin{figure*}[t]
\centering 
\includegraphics[width=0.8\textwidth]{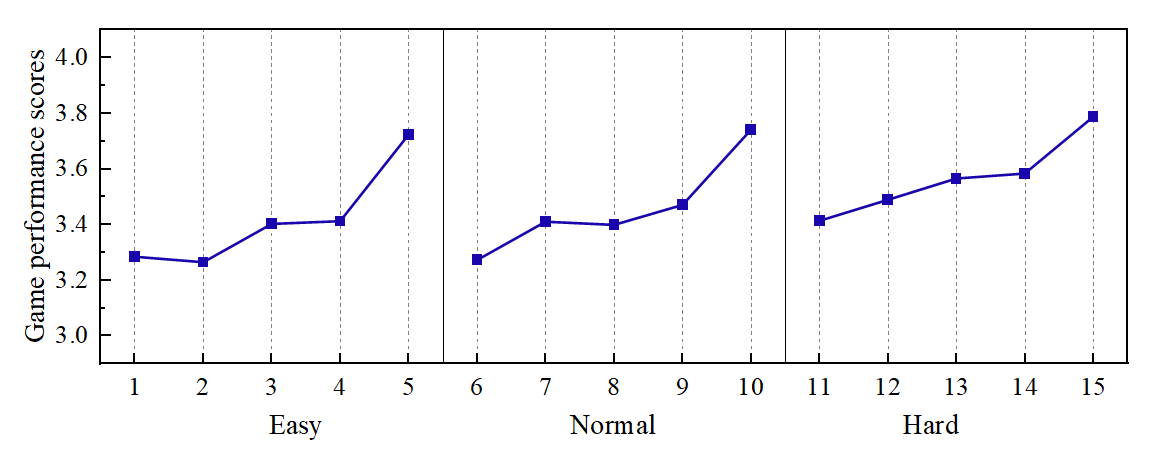}
\caption{The GS of 15 training sessions, with five sessions for easy, followed by five sessions for normal, followed by five sessions for hard.}
\label{fig:game 15}
\end{figure*}

We conducted a repeated-measures ANOVA with a 5 (NTS: N1, N2, N3, N4, N5) * 3 (game difficulty: easy, normal, hard) design to analyze the data, as shown in \autoref{fig:game 15}. The results revealed a significant main effect of NTS, indicating a substantial increase in older adults' game scores (GS) as the number of training sessions (NTS) increased $(F_{4,132}=19.455, p=0.000)$. Post-hoc Bonferroni tests indicated significant differences among the different NTSs. Specifically, the GS was significantly higher in the fifth training session (N5) compared to the first four sessions (N1-N4) (for all, $p<0.001$). As for the effect of game difficulty, there was no significant difference in the GS of older adults $(F_{2,33}=0.272, p=0.764)$ as the difficulty level increased. Furthermore, the interaction between the NTS and difficulty did not have a significant impact $(F_{8,132}=0.396, p=0.869)$.

\subsubsection{Behavioral Cognitive Tests}

\begin{itemize}

\begin{figure}[htbp]
\centering 
\includegraphics[width=\columnwidth]{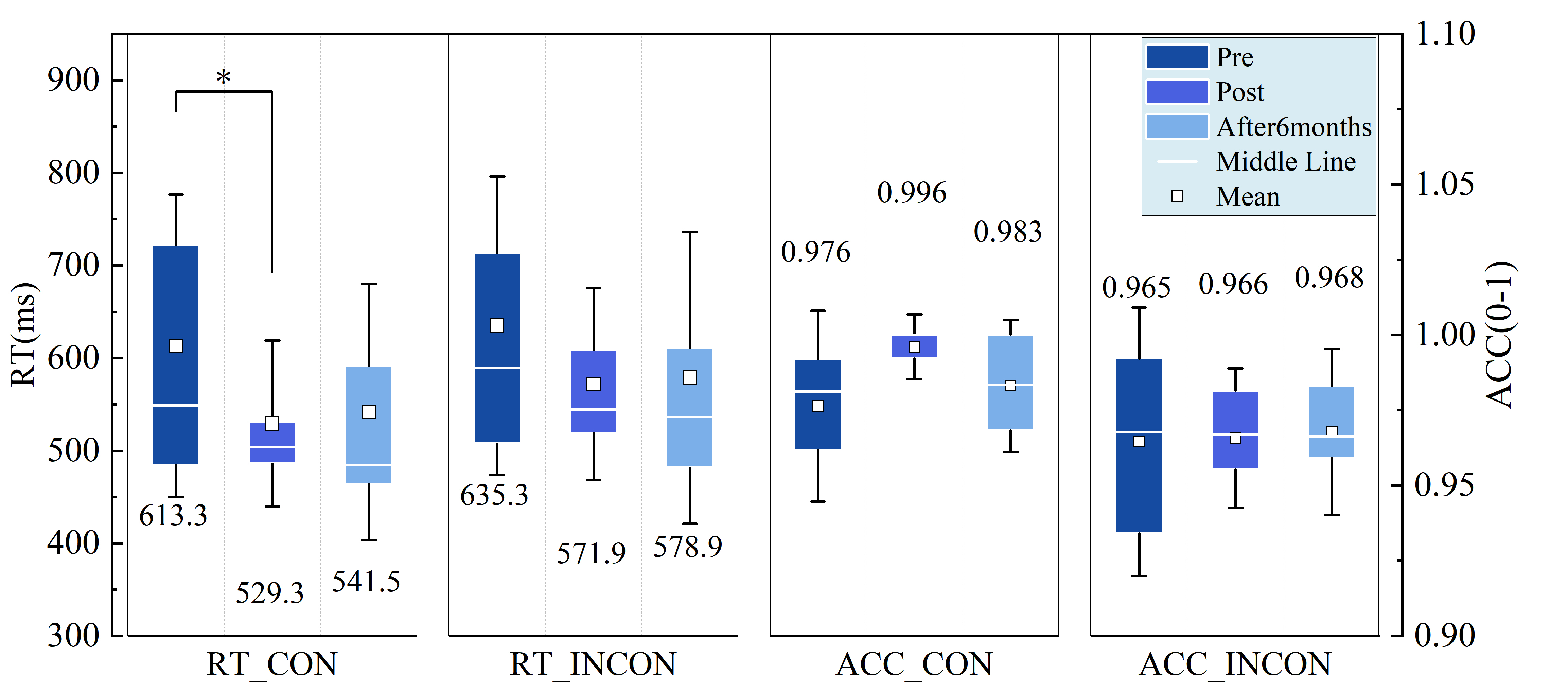}
\caption{The Stroop task's results in the pre-test, post-test, and after 6 months-test in the older adults. CON denotes that the color and word are consistent; INCON denotes that the color and word are inconsistent. RT stands for response time, reported in milliseconds; ACC stands for the accuracy of the response, which ranges from 0 to 1. ($\**: p<0.05$, $\***: p<0.01$, $\****: p<0.001$)}
\label{fig:color time}
\end{figure} 

    \item \textbf{The Stroop task:}
    We conducted a two-way repeated measures ANOVA with 3 time points (pre, post, and 6 months after training) * 2 conditions (CON, INCON) on the Stroop interference effect, as shown in \autoref{fig:color time}. The analysis of accuracy revealed no statistically significant main effect of time $ (F_{2,44}=1.993, p=0.161)$. However, a significant main effect of the condition was observed $ (F_{1,22}=15.267, p=0.001)$. Additionally, the interaction effect between time and condition did not reach statistical significance $ (F_{2,44}=1.763, p=0.192)$. Regarding the response time (RT), there was a statistically significant main effect of time $ (F_{2,44}=3.691, p=0.033)$. No significant differences were found for the main effect of condition $ (F_{1,22}=0.566, p=0.460)$. Post-hoc Bonferroni tests revealed that RT in the post-test phase was significantly shorter than in the pre-test phase $ (p=0.000)$. Furthermore, after 6 months, the RT remained significantly shorter compared to the pre-test phase $ (p=0.003)$. These findings indicate that after 20 sessions of game training, older adults exhibited significantly shorter RT, and these improvements were maintained 6 months later. The interaction between time and condition was not statistically significant $ (F_{2,44}=0.066, p=0.852)$.

\begin{figure}[htbp]
\centering 
\includegraphics[width=\columnwidth]{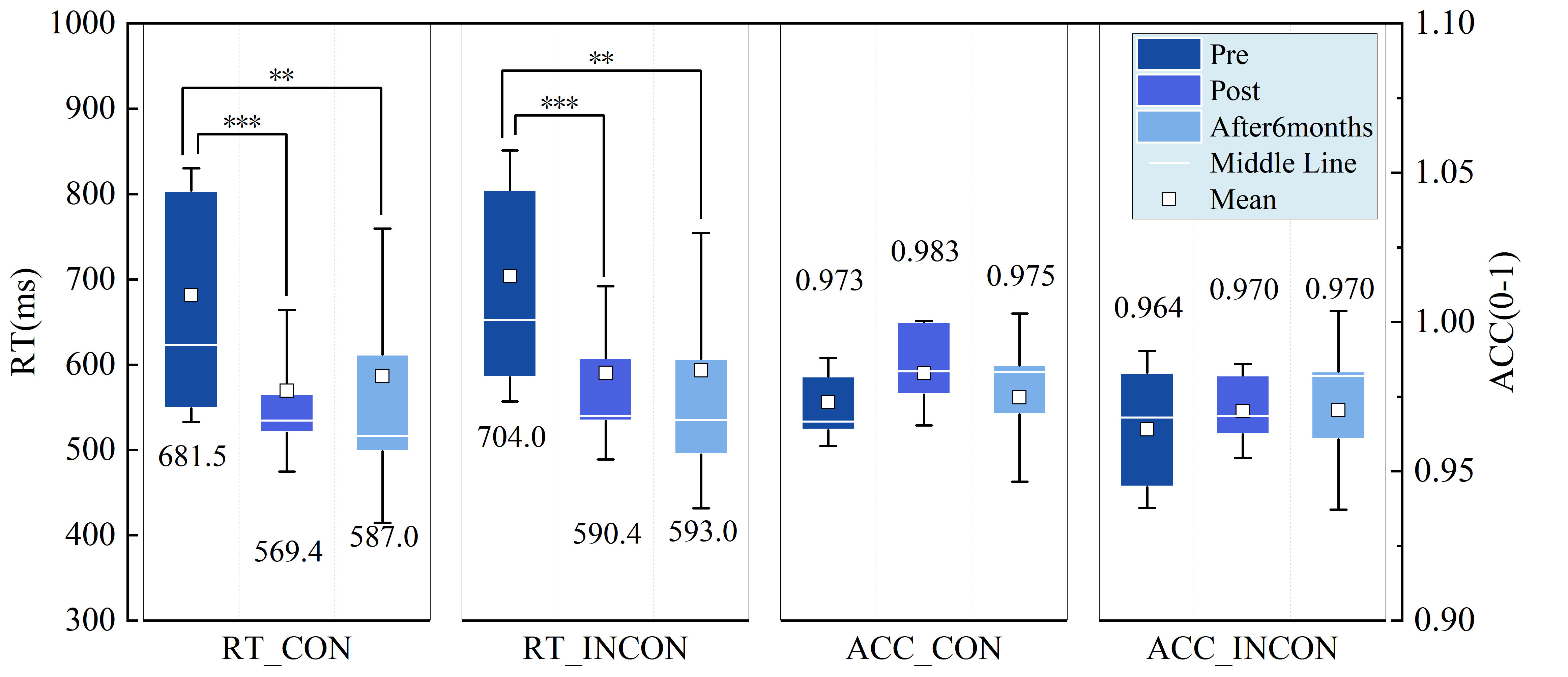}
\caption{The Reverse Stroop task's results in the pre-test, post-test, and after 6 months-test. ($\**: p<0.05$, $\***: p<0.01$, $\****: p<0.001$)}
\label{fig:meaning time}
\end{figure}

    \item \textbf{The Reverse Stroop task:}
    The analysis procedure for the Reversed Stroop task was the same as that used for the Stroop task, as shown in \autoref{fig:meaning time}. In terms of accuracy, the analysis showed no statistically significant main effect of time $ (F_{2,44}=0.661, p=0.488)$. Similarly, no significant differences were found for the main effect of condition $ (F_{1,22}=2.522, p=0.127)$. Additionally, the interaction effect between time and condition was not significant $ (F_{2,44}=0.188, p=0.777)$. In terms of response time (RT), a statistically significant main effect of time was observed $ (F_{2,44}=8.526, p=0.002)$. Post-hoc Bonferroni comparisons revealed a significant difference in RT between the pre-test and post-test phases $ (p=0.000)$. Moreover, a significant difference was found between the pre-test phase and the 6-month follow-up test $ (p=0.012)$. These findings suggest that older adults exhibited significantly shorter RT after completing 20 sessions of game training, and this improvement was maintained 6 months later. No significant differences were found for the main effect of condition $ (F_{1,22}=0.131, p=0.721)$. Furthermore, the interaction effect between time and condition was not significant $ (F_{2,44}=0.046, p=0.915)$.

    \item \textbf{The Go/NoGo task:}
    We conducted a one-way repeated measures ANOVA to analyze the scores at pre-test, post-test, and 6 months later follow-up test, as shown in  \autoref{fig:gonogo time}. The results indicated a significant effect of time on the discriminative index $d$ $ (F_{2,22}=8.195, p=0.002)$. Post-hoc comparisons revealed that d was significantly higher at post-test compared to pre-test $ (p=0.003)$, but there was no significant difference after 6 months $ (p=0.094)$. Regarding the response time (RT), a significant main effect of time was observed $ (F_{2,22}=4.933, p=0.017)$. Bonferroni post-hoc tests revealed that RT at post-test was significantly shorter than at pre-test $ (p=0.047)$. Similarly, RT after 6 months was also significantly shorter $ (p=0.032)$.

\begin{figure}[htbp]
\centering 
\includegraphics[width=0.7\columnwidth]{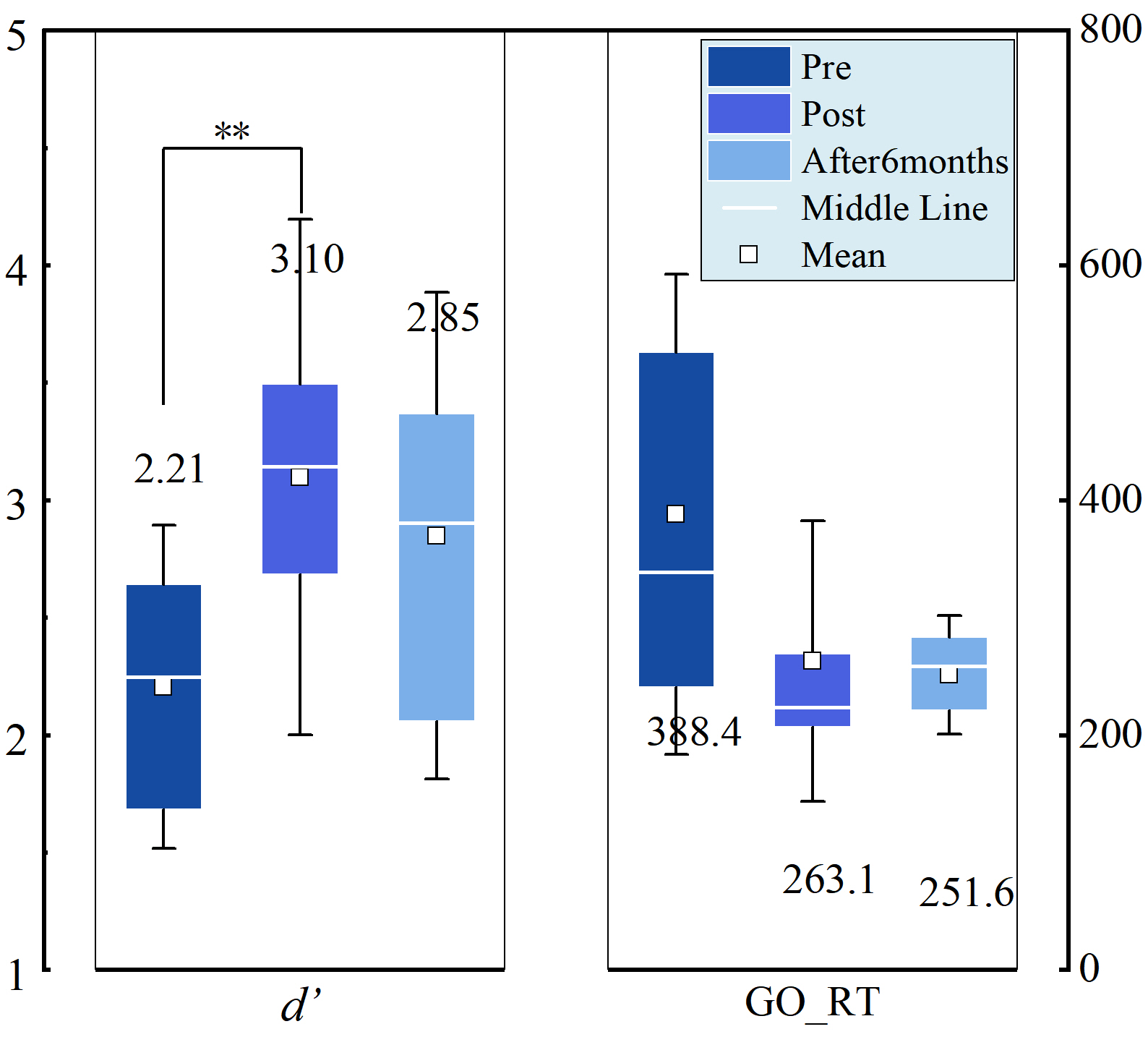}
\caption{The Go/NoGo task's results in the pre-test, post-test, and after 6 months-test in the older adults. GO RT stands for the response time of the Go stimulate, reported in milliseconds. The $d$ is the discriminability index, with a shorter RT Go, a greater ACC NoGo, and a higher d indicating superior ability. ($\**: p<0.05$, $\***: p<0.01$, $\****: p<0.001$)}
\label{fig:gonogo time}
\end{figure}

\end{itemize}

\subsection{Qualitative Feedback}

\rv{We collected qualitative data through observation and open-ended questions. Before the analysis, the data was translated into English using DeepL and was checked by the first authors for inaccuracies \cite{deepl}. We then analyzed the data using a reflexive thematic analysis approach. We grouped four deductive categories corresponding to our design goals: \textbf{Benefits from Exergame Training, Motivation and Compliance, Safety and Usability, and User Preferences}. We coded the interview data using inductive codes (e.g., ``injury risk'', ``accessibility of exercise opportunities'', ``game scene should be improved'') under these categories. Following this step, we performed an affinity mapping activity \cite{beyer1999contextual,sparkes2016qualitative} and based on this, we created the following themes.} \par
Based on the quantitative data, we first discuss insights from qualitative data to further unpack the benefits of long-term cognitive training. Then we will discuss how our design goals have worked together to improve participants' motivation and compliance.

\subsubsection{\rv{Benefits from Exergame Training}}
\
\newline
\rv{Engaging in meaningful activities is essential for the well-being of people in their later years. Participation in such activities offers older adults an avenue to cultivate a healthy lifestyle, engage in regular physical exercise, maintain positive psychological outlooks, and foster social connections \cite{zhao2023older,ryan2017self,tierney2020enjoyable}. With the progression of aging, their opportunities to engage in meaningful activities may shrink due to the decline in their health and mobility. All participants (N=12) expressed that our game transformed their meaningless daily activities to intellectual challenges, and they greatly benefited from our exergame training. Through the observation of the training process and the analysis of the workshop outcomes, the participants mainly attributed the benefits of LightSword to three factors: 

First, \textbf{we found further support consistent with the quantitative results that LightSword has potential as an innovative tool to keep cognition active in older adults.} Most participants (N=10) indicated that VR exergame training provides novel, enriching, and meaningful experiences for older adults and also keeps their brains active, improves concentration, and enhances their reaction speed through training. As one participant (P6) explained, \emph{``The VR games challenged my brain in ways I didn't expect, making my brain active in thinking and reacting fast, maybe it's good for certain cognitive domains in the brain, I am not sure. But I am sure it's better than watching TV or playing TikTok. What's more, I feel I have more focus on some activities.''} \par
Secondly, \textbf{LightSword training provided older adults an opportunity to do physical exercise.} Due to the mobility limitations of older adults and the impact of the pandemic, their physical exercise routines faced significant constraints. Most participants (N=8) agreed that LightSword training would positively affect their upper body. Some commented that \emph{``it would be good for [their] fitness''} (P2). According to one participant, especially over time, one would get better at performing swing and at the evaluated aspects, which \emph{``increases the average upper body power and also reduces the risk of injury. More importantly, I have noticed a significant stabilization in my arms, as I previously experienced some trembling.''} (P8). \par
Thirdly, \textbf{some participants (N=7) believe that LightSword training has substantial benefits in terms of positive psychology, including their confidence in using new technologies and their sense of accomplishment and social connection.} P3 noted, \emph{``Since retirement, I have no access to do some intellectually stimulating activities. My days are full of watching TV, doing housework, and taking care of my grandchildren, which leads to a negative feeling of having nothing to look forward to. However, watching my training performance improve day by day, I can exchange experiences and competitions with my friends, which gives me a strong sense of accomplishment. I like this game.''} As one participant (P9) elaborated: \emph{``Even though I am getting older, I have motivations to learn new technologies to keep up with the trends of the times and satisfy my intellectual and emotional needs. Using VR for training makes me feel like I'm at the forefront of technology, and I'm very proud.''} } \par
\sout{The participants mainly attributed the benefits of LightSword to two factors. The first factor was the accessibility of physical exercise opportunities: ``We keep swinging our arms quickly to the target height, which exercises the upper body.'' Secondly, they emphasized the cognitive training of LightSword: ``We keep thinking in this novel game, maybe it's good for certain cognitive domains in the brain.'' No one had raised the risk of this game.}

\subsubsection{\rv{Motivation and Compliance}}
\
\newline
\rv{The motivation and engagement in training activities were commonly measured by the participation rate in gaming-based exercises within extended training programs. During the 20 training sessions, none of the participants dropped out of the gaming which manifests high motivation of the older adults to continue with LightSword training. They also did not report impatience or boredom. As final feedback after the last 20th gaming session, nine older adults (75$\%$) mentioned that the time of the gaming passed too quickly, they would participate in the training again, and they would recommend the training to their friends. Four older adults (30$\%$) suggested that three months of training is better.}\par

\rv{Through data analysis of the workshop, participants identified three primary reasons why LightSword enhances users' motivation and compliance. First of all, \textbf{participants expressed their desire to continue the training because the progressively increasing difficulty of the game made the game challenging, and completing the game provided them with a sense of achievement.} As the game difficulty increased, the music rhythm accelerated, requiring participants to enhance their arm movement speed and coordination to achieve higher scores. Additionally, they also need more focus and greater effort to meet the challenges. Overcoming these higher-level challenges provided a greater sense of achievement and satisfaction, which could encourage participants to maintain prolonged participation and compliance. The game design included more than 50 songs that were familiar to older adults. Therefore, although the game training was repeated 20 times, it did not become boring. Most participants perceived (N=9) that the combination of music and movement was very interesting. P5 stated: \emph{``it is great, wonderful, music is great. I loved it here. [...] The music, something completely new, a large variety in motions.''} Another participant emphasized, \emph{``I used to enjoy playing mahjong, but playing it every day after retirement became boring. However, participating in this training is different. It feels like going through levels and fighting monsters. I look forward to the next level of difficulty and music, even though higher difficulty makes me tired.''(P7)} 

Second, \textbf{almost all participants (N=10) believed that individual records and daily champions played an important role, which could enhance social connections and boost their enthusiasm and engagement.} Older adults often face a higher risk of social isolation as retirement and their children becoming independent lead to a reduction in their day-to-day social interactions. However, older adults increasingly require social connections to obtain emotional support and alleviate loneliness. Through these connections, they can share their experiences and progress, mutually encouraging and competing with each other. This environment fosters learning and enhances engagement. For instance, P1 emphasized:  \emph{``I participate in the training sessions with my friends (P2), and sometimes we discuss our experiences. The friendly competition between us is enjoyable, reminiscent of our school days, making us feel like we're back in our childhood. haha...''} 

Third, the \textbf{feedback feature of LightSword was particularly appreciated by some participants (N=5) and mentioned as a reason for long-term use.} They expressed that this feature significantly boosted their motivation and interest, thereby stimulating more active involvement in the training sessions. As P11 explained: \emph{``Upon hitting the object, I see the real-time score, which motivates me to do better. I particularly enjoy that Excellent visual effect.''} However, some participants (N=2) argued that such real-time feedback was a distraction and could impact the focus of the training. P10 emphasized,  \emph{``Sometimes, I focus solely on the real-time score and neglect the next target, causing a distraction in my training. I think it would be better without it.''} } \par
\sout{Older adults did not report impatience or boredom during the 20 training sessions: ``The songs were totally different for each session, which brings a whole new gameplay experience.'' ``As the number of training sessions increases, the gradual increase in difficulty levels is challenging for us.'' The real-time feedback feature of the game was particularly appreciated by some participants and mentioned as a reason for long-term use: ``When we play the game, success is rewarded, failures ignored, we could get a lot of positive encouragement from the game.'' For participants, individual records and daily champions played an important role: ``I enjoy competing with my friends.'' Most older adults suggested that three months of training is acceptable.}

\subsubsection{Safety and Usability}
\
\newline
\rv{\textbf{Overall participants did not report any serious safety and usability issues while using LightSword.} In terms of safety, many participants (N=11) consistently expressed that the game provided a secure and stable experience because this game allowed them to engage in the training while seated. One participant highlighted the positive side of this game by comparing it to another commercial VR game: \emph{``I can sit to play game, which supports me a sense of security, [...], While playing BeatSaber \cite{games2019beat}, I usually stand, but I'm always concerned about the risk of falling or bumping into objects in the real environment and caused injured. I'm scared to walk in the virtual world. It makes me uneasy.''} (P3) Regarding usability, participants gave positive feedback on the game's ease of use and operation. They thought LightSword included rich movements and could offer valuable options (being mobile and social value). One participant mentioned, 
\emph{``The game provides excellent guidance, making it accessible even to those who are unfamiliar with virtual reality. Participants can easily learn to raise their arms and work with their trunks, with breaks between these actions to avoid sustained muscle tension. This design prevents excessive strain during prolonged gameplay. Overall, the game is highly enjoyable and user-friendly.''} (P12)

However, \textbf{there were some concerns reported by participants and associated LightSword with some drawbacks.} One participant reported the possibility of losing physical world awareness while playing this game: \emph{``I couldn't remember where I was in the room when I was in VR.''} (P4) Two participants exhibited mild discomfort following one training session (not the first). Notably, one participant reported some usability problems due to the technical VR setup: \emph{``Problems with sharp vision, which strained the eyes. [...] The cable around me was disturbing.''} (P9)} \par

\sout{Overall LightSword did not cause serious safety and usability issues. One participant highlighted the positive side of this game by comparing it to another commercial alternative: ``We can sit to play game, less movement means safety for participants than in other VR games like Beat Saber \cite{games2019beat}, in which I'm afraid I'll fall down.'' However, some participants were confused about their physical location: ``I couldn't remember where I was in the room when I was in VR'.'' Only two participants exhibited mild discomfort following one training session (not the first). Notably, older adults reported some usability problems due to the technical VR setup: first, ``Problems with sharp vision, which strained the eyes.'' Second, ``The cable around me was disturbing.''}

\subsubsection{User Preferences}
\
\newline
\rv{Participants expressed a preference for some features, they also suggested some improvements for LightSword could be considered, like establishing VR environments that correspond to the songs, providing proper feedback, and setting variations of game levels. A few participants preferred \emph{``change nothing''} (P2,6) Others emphasized the parts they liked the most, like \emph{``[...] definitely keep the music and Daily Champion.''} (P1) \emph{``[...] Sitting and playing with fewer head movements were worth remaining.''} (P11) Still, we received many ideas on how to improve LightSword. Some participants proposed improvements to the game world, such as \emph{``[...] synchronizing the game scenes with the song's changes, which would create a really cool experience.''} (P10), and \emph{``a more accurate virtual sword would be desirable to play better.''}(P4) A few wished to have more feedback: \emph{``When my hit point deviates to the left or right, I wish to receive corresponding feedback. Additionally, the Excellent bubble and sound appear simultaneously after achieving five successful GREAT, it's better.''} (P5) A multiplayer mode to facilitate social interaction was also suggested for improvement: \emph{``I think it would be better if we could participate in the training as a team, singing and doing actions simultaneously.''} (P12)} \par
\sout{Older adults suggested improvements for LightSword to the game scenes and variation of game levels. ``You could consider synchronizing the game scenes with the song's changes, which would create a really cool experience.'' There were still some designs that were worth keeping. ``Sitting and playing with fewer head movements were worth remaining.''}

\section{Discussion}

The main goal of this research was to train cognitive inhibition for the long term in older adults, so we designed a customized VR exergame to fit their needs. We considered both prior literature and the feedback of older adults to design LightSword. Then, we evaluated our prototype through the user study. In this section, we discussed the findings focused on the research questions and provided design implications for future research.

\subsection{RQ1: LightSword enhanced cognitive inhibition in older adults both in short-term effects and long-term effects}

Our results showed older adults performed better in the post-test and 6-month follow-up test compared to the pre-test, both in terms of game scores and behavioral cognitive tests. This finding showed that LightSword can improve cognitive inhibition from the short-term and long-term effects. The significant improvements may be related to the mechanisms of the Stroop paradigm. The Stroop effect effectively engaged selective attention in older adults through the interference of colors and words, allowing them to focus on salient stimuli while inhibiting and filtering out irrelevant ones. This cognitive challenge stimulated the anterior cingulate cortex and the dorsolateral prefrontal cortex, which were crucial in decision-making, attentional control, and inhibitory functions. Furthermore, the dual-task paradigm required older adults to simultaneously conduct multiple tasks under limited cognitive resources. This training could help participants enhance their efficiency in allocating attention, memory, and task execution, ultimately leading to improved multitasking abilities. The alternate-runs paradigm demanded older adults to rapidly switch between different tasks. This training could enhance cognitive flexibility and switching skills, including the ability to suppress distractions and maintain concentration between tasks. By naturally integrating these three paradigms into game design, individuals were exposed to more complex and diversified cognitive inhibition challenges, ultimately better equipping them to meet the cognitive demands of real-life situations. \par

Surprisingly, the benefits of the training also extended to untrained Go/NoGo task, which was not consistent with the brain regions stimulated by the three cognitive activation paradigms above. Although Go/NoGo task and Stroop task were related to cognitive inhibition, they had different attentional focus and execution requirements. One possible reason was that the combination of three paradigms requires coordinated efforts among different brain regions to process conflicting information. This may contribute to strengthening the connection between the anterior cingulate cortex and the dorsolateral prefrontal cortex with other cognitive regions, thereby enhancing the capacity for response inhibition. These findings were consistent with the Wang et al. \cite{75}, but inconsistent with the results of Zhao et al., who argued that transfer effects of cognitive inhibition only occur in children and are not demonstrated in adults \cite{74}. However, further experimental validation was required to confirm whether this combination enhances the training effect. \par

What is inconsistent with our hypothesis was that we did not observe a significant increase in the accuracy, both in the Stroop test and in the Reversed Stroop test. We suspected one possible reason was that we allocated excessive time to older adults in the behavioral cognitive tests, giving them ample time (3000ms) for judgment. As a result, participants demonstrated very high accuracy levels in the pre-test, which limited the degree of improvement. Another aspect worth noting is that older adults' game GREAT rate did not exhibit a declining trend with increasing difficulty but instead showed a gradual increase. Achieving a GREAT score for each block was full of challenges. Older adults must not only accurately judge the blocks but also pay attention to the speed and height of the swing arm. This result implied that after training, upper limb movements in older adults improved and were not adversely affected by increased difficulty. However, we emphasized the necessity for more extensive and focused studies to demonstrate the physical improvement in older adults. 

\subsection{RQ2: Older adults exhibited a positive attitude toward long-term training with LightSword}

LightSword was implemented based on older adults' feedback. Therefore, we attributed the high appreciation by older adults to the targeted design process of our game. Aligning with previous studies, we showed that LightSword can lead to positive participants' experiences \cite{wei2023bridging, 41, chao2013effects, stanmore2017effect, yoo2016vrun}. So we extended the results of these studies into cognitive inhibition. Almost all older adults expressed a positive attitude toward long-term training with LightSword. And they never showed impatience or boredom during the 20 training sessions, because the songs were totally different for each session. The older adults remained engaged and focused throughout the training. Because VR can provide a strong sense of presence, which directs the user's attention to the training \cite{38,dilanchian2021pilot}. \rv{Furthermore, the dynamically increasing game difficulty presented challenges for older adults, helping to maintain their focus.} Older adults were competing to be daily champions because they considered it a reward and an honor, bringing them a sense of accomplishment. Only two older adults exhibited mild cybersickness following one training session (not the first session), which may have been attributed to their physical condition on that particular day. However, some older adults criticized problems with sharp vision strained the eyes, which was uncomfortable and could potentially lead to vision degradation. The participants also reported some usability problems due to the technical VR setup like \emph{``The cable around them was disturbing''}. These were issues we need to make further improvements. \par

While the majority of older adults expressed their intention to continue using LightSword, it was important to note that this feedback alone cannot serve as conclusive evidence for long-term training. However, we emphasized the necessity for more extensive and focused studies. Additionally, older adults had provided specific improvement suggestions, including a more dynamic virtual environment, the complexity of difficulty levels, and the possibility of matching scenes and levels to enhance gameplay variety. \sout{The last point held particular significance for the sustained success of future cognitive VR exergames.} To ensure long-term training, developers should focus on varied gameplay with high usability. Incorporating rhythmic elements, as seen in our game and popular commercial games like Beat Saber, can be extended by introducing new songs and mappings. However, it remained an open research question on how to effectively enhance compliance for advanced training routines, particularly in long-term training.

\subsection{Design Implications}
\rv{In the section, we aim to discuss how our study has inspired design implications within the relevant field and how these insights can guide future research. Concurrently, we explore the limitations of our study and the potential direction for future research improvements.}

\subsubsection{Adapt the Difficulty According to Participants' Fitness Level and Improvements}
\
\newline
We emphasized the importance of dynamically increasing game difficulty, a key feature in ensuring motivation and compliance in long-term training \cite{mueller2016exertion, xu2020results, 81, huber2021dynamic}. In the design process, our game assessed the participants' cognitive inhibition condition (reaction time) and then customized the difficulty based on the maximum, minimum, and average values. This design enabled our game to provide challenges for every older adult, irrespective of their fitness level and ability, without overwhelming beginners. In the pre-test, there were significant differences in game scores at three difficulty levels. The Cor rates for all levels were well within our anticipated window of 60$\%$ to 90$\%$, the GREAT rates were well within our anticipated window of 30$\%$ to 70$\%$. Hence, we could assume that our three difficulty settings worked well in aligning the exergame's challenge to the individual abilities. \rv{Our research findings were in line with Anguera's positive attitude towards the adaptive staircase difficulty, which was lacking in many studies on game design \cite{26,44,gizatdinova2022pigscape, kruse2021long}. Our findings suggested it is a potential direction that assessing the participant's performance and setting dynamic increasing difficulty in future research. However, there are some open challenges in the real-time performance assessment of participants. Researchers need to explore more innovative algorithms to address this issue in future work.} \sout{In future research, assessing the individual participant's ability and setting dynamic increasing difficulty was important. Furthermore, incorporating adaptive algorithms into the game design to continuously monitor participants' performance and switch difficulty on demand was a potential solution. However, in our game, we abandoned adaptive difficulty algorithms because participants' experience is more dependent on the songs, and excessive changes could lead to confusion, especially among older adults.}

\subsubsection{Incorporate Music and Rhythmic Elements in VR Exergame}
\
\newline
  Similar to other commercial exergames like Beat Saber \cite{games2019beat}, we aligned the participants' movements to the beats of songs. This game design was widely appreciated by the participants who stated \emph{``[...] definitely like the music and movements''.} \rv{Our results indicated that incorporating music and rhythmic elements into VR exergames could better motivate older adults to play exergames with higher engagement. This was inconsistent with the research findings of Boot et al. \cite{boot2013video}, who claimed that older adults perceive action games as lacking in both enjoyment and cognitive benefits. However, movements synchronized with music rhythms may engage older adults in excessive exercise, potentially leading to exhaustion. Therefore, the challenge lay in finding a balance between the music's rhythm and the frequency of movement, which remained an open question.} \sout{However, it was essential to limit the break and interval between two elements, as fast swing and arms exercises are exhausting for older adults. In our game, we incorporated resting passages to allow participants to relax and take a break. Furthermore, music therapy has become increasingly accepted in cognitive interventions in recent years, especially when selecting music linked to an individual's biography. Music therapy can evoke memories and combat depression. It was worth noting that we speculated that multimodal integration of visual and auditory stimuli in training could enhance the effect of training. However, further evidence was needed to substantiate this hypothesis.} 
 
 \subsubsection{Safety Concerns in VR Exergames}
\
\newline
Older adults were particularly susceptible to experiencing cybersickness and discomfort during VR exergames due to age-related declines in their balance and proprioceptive capabilities, reduced adaptability to physical exertion, and limited familiarity with VR devices. Moreover, certain strenuous physical activities in VR exergames, such as jumping, dodging, or rapid movements, could pose physical challenges for older adults, thereby increasing the risk of falls or injuries. That was a major reason why we kept older adults seated in the game. Additionally, movements like forward jumping or walking could quickly transport users to the borders or even beyond the play space, which required ample room for the participant to avoid potential harm. This consideration is especially crucial for game developers who must accommodate various consumer play spaces, which may not always be generously sized.

\section{Limitations and future work}

Our study identified several limitations that should be acknowledged and some future works. Firstly, the generalizability of the findings is limited due to experimental constraints and a small sample size. We specifically focused on evaluating cognitive improvements in healthy older adults after 20 game training sessions, and it remained uncertain whether the game would bring similar benefits to individuals in other age groups. Another limitation was that our experiment mainly included healthy older women due to pandemic-related recruitment challenges. Although previous research reported no significant gender differences in game training outcomes \cite{17}, we must acknowledge that our sample primarily consisted of female older adults. \rv{Furthermore, we could explore the cognitive differences among older adults of different age groups (younger, middle-aged, and older) in future research.} Additionally, further experimental validation was required to confirm whether the combination of three cognitive activation paradigms enhances the training effect. Lastly, our evaluation of cognitive abilities was solely based on game performance scores and behavioral test indicators. However, previous research has demonstrated that objective physiological measures, such as EEG and heart rate, can also serve as indicators of cognitive abilities \cite{35}. For example, Dennison identified strong correlations between physiological data (including EEG waves, blink frequency, and heart rate) and scores on the Simulator Disease Questionnaire, suggesting that physiological characteristics can predict cognitive function \cite{dennison2016use, vanneste2021towards}. In future studies, the integration of physiological and behavioral indicators may offer a more comprehensive assessment of cognitive performance.

\section{Conclusion}

In this work, we designed the LightSword exergame which combined physical and cognitive exercises to train older adults' cognitive inhibition for the long-term. Targeted improvements were considered in the game design in order to meet the needs of long-term cognitive training for older adults: (1) Select appropriate cognitive activation paradigms to train cognitive inhibition; (2) Customized settings to enhance motivation in encouraging active engagement; (3) Maintain freshness by updating songs to improve compliance and support long-term training; (4) Reduce head movement and remain seated to ensure safety in prolonged physical exercise. Then, an eight-month longitudinal user study with 12 older adults aged 60 years and above was conducted to demonstrate the effectiveness of LightSword in improving cognitive inhibition. The results showed older adults' cognitive abilities improve significantly after training, with gains persisting for 6 months. To our knowledge, we were the first to use a custom-designed VR exergame for long-term cognitive inhibition training for older adults.

\begin{acks}
This work was partially supported by a grant from the Beijing Outstanding Young Scientist Program (BJJWZYJH01201910048035), partially supported by 2023 Guangzhou Science and Technology Program City-University Joint Funding Project (No. 2023A03J0001) and Guangdong Provincial Key Lab of Integrated Communication, Sensing and Computation for Ubiquitous Internet of Things (No. 2023B1212010007).
\end{acks}

\bibliographystyle{ACM-Reference-Format}
\bibliography{ref}










\end{document}